\documentclass[11pt]{article}
\usepackage{palatino}
\usepackage{url}
\usepackage{hyperref}
\usepackage{latexsym}
\usepackage{graphicx}
\usepackage{multirow}
\usepackage[usenames,dvipsnames]{color}

\usepackage{cite}

\usepackage{pdflscape}

\title{A Comparison of Microbial Genome Web Portals}

\author{Peter D. Karp,$^1$ Natalia Ivanova,$^2$ Markus Krummenacker,$^1$ Nikos Kyrpides,$^2$  \\
Mario Latendresse,$^1$ Peter Midford,$^1$ Wai Kit Ong,$^1$ Suzanne Paley,$^1$ and Rekha Seshadri$^2$\\
$^1$Bioinformatics Research Group, SRI International, Menlo Park, USA\\
$^2$DOE Joint Genome Institute, Walnut Creek, USA\\
pkarp@ai.sri.com\\
}

\date{}

\begin{document}

\maketitle

\section*{Abstract}

Microbial genome web portals have a broad range of capabilities that
address a number of information-finding and analysis needs for
scientists.  This article compares the capabilities of the major
microbial genome web portals to aid researchers in determining
which portal(s) are best suited to solving their information-finding and
analytical needs.
We assessed both the bioinformatics tools and the data
content of BioCyc, KEGG, Ensembl Bacteria, KBase, IMG, and PATRIC.
For each portal,
our assessment compared and tallied the available capabilities.  The
strengths of BioCyc include its genomic and metabolic tools, 
multi-search capabilities, table-based
analysis tools, regulatory network tools and data, omics data
analysis tools, breadth of data content, and large amount
of curated data.
The strengths of KEGG include its genomic and metabolic tools.
The strengths of Ensembl Bacteria include its genomic tools and large number of genomes.
The strengths of KBase include its genomic tools and metabolic models.
The strengths of IMG include its genomic tools, multi-search
capabilities, large number of genomes, table-based analysis
tools, and breadth of data content.
The strengths of PATRIC include its large number of genomes, 
table-based analysis tools, metabolic models, and breadth of
data content.

\section{Introduction}

A number of web portals exist for providing the scientific community
with access to the thousands of microbial genomes that have been
sequenced to date.  This article compares the capabilities of the major
microbial genome web portals to aid researchers in determining
which portal(s) best serve their information-finding and
analytical needs.

The power that a genome web portal provides to its users is a function
of what data the portal contains, and of the types of software tools
the portal provides to users for querying, visualizing, and analyzing
the data.  Query tools enable researchers to find what they are looking
for.  Visualization tools speed the understanding of the information that
is found.  Analysis tools enable extraction of new relationships from
the data.

We assess the data content of each portal both according to the types of
data it provides (e.g., does it provide regulatory network
information, protein localization data, or Gene Ontology
annotations?), and according to the number of genomes it provides.  We
assess the software tools provided by each portal in several major
areas: genomics tools, metabolic tools, advanced search and analysis
tools, web services, table-based analysis, and user accounts.  Omics data analysis
capabilities are also assessed, but are distributed among the
preceding areas.  In each area, we enumerate
multiple software capabilities, such as the ability to paint omics data
onto pathway diagrams.  We must emphasize that many of the portals
include a significant number of other capabilities that are not within
the purview of this study.

Search tools are a particularly important part of a portal because
they determine the user's ability to find information of interest;
therefore, we provide detailed comparisons of the search tools that each
portal provides for finding genes, proteins, DNA and RNA sites,
metabolites, and pathways.  We call these multi-search tools because
they enable the user to search multiple database (DB) fields in combination.

Although user friendliness is a critical aspect of any
website, it is extremely difficult to assess objectively.  We have
assessed a small number of relatively objective user friendliness
criteria, such as the types of user documentation available, the
presence of explanatory tooltips (information windows that appear when
the user hovers over regions of the screen), and the speed of the
site's gene page.

Our criteria for inclusion in the comparison were portals with a
perceived high level of usage, large number of genomes, a relatively
rich collection of tools, and sites that are actively maintained and developed.
The portals we compare are BioCyc \cite{MetaCycNAR18} (version 22.0, April 2018),
KEGG \cite{KEGG17} (version 87.1, August 2018),
Ensembl Bacteria \cite{Kersey18} (Release 40, July 2018),
KBase \cite{KBASE18} (versions during August 2018 to October 2018),
IMG \cite{IMG17} (version 5.0 August 2018),
and PATRIC  \cite{Wattam14} (version 3.5.21, July 2018).

Related portals that are not included in this comparison are Entrez
Genomes (whose capabilities are similar to Ensembl Bacteria),
MicroScope \cite{Vallenet17} (which uses Pathway Tools for its
metabolic component and therefore has the same functionality as
BioCyc), ModelSEED \cite{Henry10} (which is a metabolic model portal,
not a genome portal), the SEED \cite{Overbeek14} (which has been
inactive for a number of years and was subsumed by the PATRIC
project), MicrobesOnline \cite{Dehal10}, iMicrobe
(\href{https://www.imicrobe.us/}{https://www.imicrobe.us/}), which is
a portal for metagenomes and transcriptomes, not for single genomes),
and Microme (\href{http://www.microme.eu/}{http://www.microme.eu/},
the Microme website largely shut down as of January 2018).

%%%% Wai Kit: KEGG.  Mario/JGI: KBase.  PeterM and Suzanne: PATRIC.  

\subsection{Summary of the Portals}

Here we introduce each portal.  Note that some portals have
some capabilities that are not covered in this comparison.
For each portal we provide a hyperlink to a sample gene page.

\subsubsection*{BioCyc}

BioCyc \cite{BioCyc17,MetaCycNAR16} is a microbial genome web portal that integrates sequenced
genomes with curated information from the biological literature,
with information imported from other biological DBs, and with
computational inferences.  BioCyc data include metabolic pathways,
regulatory networks, and gene essentiality data.  BioCyc provides
extensive query and visualization tools, as well as tools for omics
data analysis, metabolic path searching, and for running metabolic
models.  We omit discussion of many BioCyc comparative genomics and
metabolic operations under its Analysis $\rightarrow$ Comparative
Analysis menu.  Scientists can use the Pathway Tools software
associated with BioCyc to perform metabolic reconstructions and create
BioCyc-like DBs for in-house genome data.

BioCyc contains information curated from 89,500 publications.
The curated information includes experimentally determined gene
functions and Gene Ontology terms,
experimentally studied metabolic pathways, and experimentally
determined parameters such as enzyme kinetics data and enzyme
activators and inhibitors.  Curated information also includes textual
mini-reviews that summarize information about genes, pathways, and
regulation, with citations to the primary literature.  The large
amount of curated information within BioCyc is unique with respect to
other genome portals.
\ \\  
\noindent Home page: \href{https://biocyc.org/}{https://biocyc.org/}\\
Sample gene page: \href{https://biocyc.org/gene?orgid=ECOLI&id=EG10823}{https://biocyc.org/gene?orgid=ECOLI\&id=EG10823}.

\subsubsection*{KEGG} 

The Kyoto Encyclopedia of Genes and Genomes is a resource for
understanding high-level functions of a biological system from
molecular-level information.  It includes a focus on data
relevant for biomedical research (e.g., KEGG DISEASE and KEGG DRUG
databases) and includes tools for analysis of
large-scale molecular datasets generated by high-throughput
experimental technologies.
\ \\  
\noindent Home page: \href{https://www.kegg.jp/}{https://www.kegg.jp/}\\
Sample gene page: \href{https://www.kegg.jp/dbget-bin/www_bget?eco:b2699}{https://www.kegg.jp/dbget-bin/www\_bget?eco:b2699}.

%%%%\subsubsection*{ModelSEED}
%%%%
%%%%ModelSEED \cite{MODELSEED}is primarily a tool for constructing and
%%%%analyzing metabolic models.  It provides access to 40 detailed models
%%%%of Plant taxa, and also includes a collection of models of microbial
%%%%metabolism.  New models can be built from either user supplied FASTA
%%%%files or from genomes in PATRIC. These models support flux balance
%%%%analysis, gap filling, and ...

\subsection*{Ensembl Bacteria}
Ensembl Bacteria is a portal for bacterial and archaeal genomes. It does
not have any data or tools for metabolism, pathways or compounds, focusing on genes
and proteins.  Its strengths seem to be in its large collection of gene and protein
family data. Its capabilities are somewhat different from other
Ensembl sites.  In addition to BLAST, it includes a hidden Markov
model (HM) search tool for protein motifs.  Pan-taxonomic comparative tools are available for key species.  It also includes Ensembl's variant effect predictor, which can predict functional consequences of sequence variants.
\ \\  
\noindent Home page: \href{https://bacteria.ensembl.org/}{https://bacteria.ensembl.org/}\\
Sample gene page: \href{https://bacteria.ensembl.org/Escherichia_coli_str_k_12_substr_mg1655/Gene/Summary?g=b2699;r=Chromosome:2822708-2823769;t=AAC75741;db=core}{https://bacteria.ensembl.org/Escherichia\_coli\_str\_k\_12\_substr\_mg1655/Gene/Summary?g=b2699;r=Chromosome:2822708-2823769;t=AAC75741;db=core}.

\subsubsection*{KBase}

KBase is an environment for systems biology research that provides
more than 160 applications to support user-driven analysis of a
variety of data ranging from raw reads to fully assembled and
annotated genomes, and metabolic models.  In addition to its
genome-portal capabilities, KBase \cite{KBASE16} enables users to
assemble and annotate genomes, to analyze transcriptomics data, and to
create metabolic models for organisms with sequenced genomes.  Once a
model is created, it can be analyzed using phylogenetic, expression
analysis, and comparative tools. KBase also allows users to integrate
custom code into their analysis pipeline and enables addition of
external applications by their developers using a software development
kit (SDK). Its other major aim is to support
reproducible computational experiments, on models, that can be
published and shared with other users.
\ \\
\noindent Home page: \href{https://kbase.us/}{https://kbase.us/}\\
Sample gene page: \href{https://narrative.kbase.us/#dataview/35926/2/1?sub=Feature&subid=b2699}{https://narrative.kbase.us/\#dataview/35926/2/1?sub=Feature\&subid=b2699}.

\subsubsection*{IMG}

The Integrated Microbial Genomes (IMG) system is a resource for
annotation and analysis of sequence data, integrated with
environmental and other metadata to support genome and microbiome
comparisons. In addition to being the vehicle for release of the data
generated by the DOE Joint Genome Institute, it provides a suite of
analytical and visualization tools available to explore and mine the
data for biological inference. Custom data marts dedicated to specific
research topics like synthesis of secondary metabolite (IMG-ABC) or
viral eco-genomics (IMG/VR), are also included. Users can submit their
own data and metadata for integration in the system.
  \ \\
\noindent Home page: \href{https://img.jgi.doe.gov/}{https://img.jgi.doe.gov/}\\
Sample gene page: \href{https://img.jgi.doe.gov/cgi-bin/m/main.cgi?section=GeneDetail&page=geneDetail&gene_oid=646314661}{https://img.jgi.doe.gov/cgi-bin/m/main.cgi?section=GeneDetail\&page=geneDetail\&gene\_oid=646314661}.

\subsubsection*{PATRIC}

PATRIC is designed to support the biomedical research community's work
on bacterial infectious diseases via integration of vital pathogen
information with data and analysis tools. Data is integrated across
sources, data types, molecular entities, and organisms. Data types
include genomics, transcriptomics, protein-protein interactions, 3D
protein structures, sequence typing data, and metadata.  It supports both genome
assembly and annotation (RAST), and RNA-seq data analysis via a job submission
system.
\ \\  
\noindent Home page: \href{https://www.patricbrc.org/}{https://www.patricbrc.org/}\\
Sample gene page: \href{https://www.patricbrc.org/view/Feature/PATRIC.511145.12.NC_000913.CDS.2820730.2821791.rev}{https://www.patricbrc.org/view/Feature/PATRIC.511145.12.NC\_000913.CDS.2820730.2821791.rev}.

\section{Results}

We assessed the software and data content capabilities of each portal
according to a number of topic areas, such as genomics-related tools
and metabolism-related tools.  We chose topic areas that we considered
to be core elements of a microbial genome information portal --- that
is, a web site that counts among its primary missions providing users
with data and knowledge regarding sequenced microbial genomes.  A
number of the portals contain functionality outside of that mission,
for example, some portals contain software tools for annotating
microbial genomes (e.g., performing assembly and gene-function
prediction).  We did not include such functionality because we
considered it outside the scope of a microbial genome information
portal.  In many cases, we added new criteria within a topic area
(meaning rows within our comparison tables) as we learned about each
portal, such as adding the ability of Ensembl Bacteria to predict the effects
of sequence variants.  Our choice of criteria is validated by the fact
that many of the criteria are shared among some or many of the portals.

For several of the topic areas, we provide multiple
tables to assess software capabilities, with one or two tables
focusing on DB search capabilities and another table focusing
on other capabilities in that area.
For example, Tables~\ref{tab:gene-protein-multi-search} and
\ref{tab:site-multi-search} describe genomics multi-search tools, and
Table~\ref{tab:genomics-tools} describe other genomics software tools.

We attempted to be as diligent as possible when evaluating each
portal's capabilities, however, being non-expert navigators of KEGG,
Ensembl Bacteria, KBase, and PATRIC, we may have overlooked or
misjudged some element of those portals.

\begin{landscape}

\begin{table}[!h]
{\small
\centerline{
\begin{tabular}{|l|c|c|c|c|c|c|c|} \hline
{\bf Tool}                 & {\bf BioCyc} & {\bf KEGG} & {\bf Ensembl Bacteria} & {\bf KBase} & {\bf IMG} & {\bf PATRIC} \\ \hline \hline
Genome Browser                        & YES & YES & YES   & YES & YES & YES \\ \hline
-- Operons, Promoters, TF binding sites &YES& no  & no    & no  & partial   &  YES   \\ \hline
-- Depicts Nucleotide Sequence        & YES & YES & YES   & YES & YES   & YES    \\ \hline
-- Customizable Tracks                & YES & no  & YES   & no  & partial  & YES \\ \hline
-- Comparative, by Orthologs          & YES & no\footnotemark[1]  & no   & no  & YES & YES \\ \hline
-- Genome Poster                      & YES & no  & no   & no   & no  & no \\ \hline
Retrieve Gene Sequence                & YES & YES & YES  & YES  & YES & YES \\ \hline
Retrieve Replicon Sequence            & YES & YES & YES  & no  & YES & YES \\ \hline
Retrieve Protein Sequence             & YES & YES & YES  & YES  & YES & YES \\ \hline
Nucleotide Sequence Alignment Viewer  & YES & YES & no   & no  & YES & YES \\ \hline
Protein Sequence Alignment Viewer     & YES & YES & no   & no  & YES & YES \\ \hline
Protein Phylogenetic Tree Analysis    & no  & YES & no   & YES  & YES & YES  \\ \hline
Sequence Searching by BLAST           & YES & YES & YES  & YES & YES & YES \\ \hline
Sequence Pattern Search               & YES & YES & no   & YES & YES & no  \\ \hline
Sequence Cassette Search              & no  & YES & YES  & YES & YES & no  \\ \hline
Orthologs                             & YES & YES & no   & YES & YES & YES \\ \hline
Gene/Protein Page                     & YES & YES & YES  & YES & YES & YES \\ \hline
Enrichment Analysis (GO Terms)        & YES & no  & no    & YES & no  & no  \\ \hline
Enrichment Analysis (Regulation)      & YES & no  & no    & no  & no  & no  \\ \hline
Omics Dashboard                       & YES & no  & no    & no  & no  & no  \\ \hline
Multi-Organism Comparative Analysis   & YES & YES & YES  & YES  & YES & YES \\ \hline
Horizontal Gene Transfer Prediction   & no  & no  & no   & no  & YES & no  \\ \hline
Fused Protein Prediction              & no  & no  & no   & no  & YES & no  \\ \hline
Alternative ORF View                  & no  & no  & no   & no  & YES  & YES \\ \hline
Genome Multi-Search                   & YES & no  & no  & no  & YES & YES  \\ \hline
gANI Computations                     & no  & no  & no  & YES  & YES & YES \\ \hline
Kmer Frequency Analysis               & no  & no  & no  & no  & YES & no \\ \hline
Synteny Comparison                    & no  & no  & no  & YES  & YES &  no \\ \hline
Proteome Comparisons                  & YES & no  & no  & YES  & YES & YES  \\ \hline
Statistical Analysis, Genome          & YES & no  & no  & no   & YES & no  \\ \hline
Statistical Analysis, Expression      & no  & no  & no  & YES  & YES & YES \\ \hline
Genome Function Comparison            & no  & no  & no  & YES  & YES & YES \\ \hline
Insert Genomes into Reference Trees   & no  & no  & no  & YES  & no & YES\footnotemark[2]  \\ \hline
Predict Effects of Sequence Variants  & no  & no  & YES & no  & no  & YES \\ \hline
%% Totals                                22   14    11      18    27    23
\end{tabular}
} }
\caption{\label{tab:genomics-tools}
{\bf Genomics Tools Comparison.}  ``Partial'' means that
  the tool provides some but not all of the indicated functionality.
$^{1}$KEGG does have a rudimentary tool for this purpose, but it is
  not based on a zoomable genome browser.  $^{2}$PATRIC
  supports construction of trees from an arbitrary set of in-group and out-group genomes.
}
\end{table}

\begin{table}[!h]
\centerline{
\begin{tabular}{|l|c|c|c|c|c|c|c|} \hline
{\bf Tool} & {\bf BioCyc} & {\bf KEGG} & {\bf Ensembl Bacteria} & {\bf KBase} & {\bf IMG} & {\bf PATRIC} \\ \hline \hline
Gene Name          & YES  & YES & YES    &  YES  & YES & YES    \\ \hline
Product Name       & YES  & YES & YES    &  YES  & YES & YES    \\ \hline
Database Identifier & YES & YES & YES    &  YES  & YES & YES    \\ \hline
EC Number          & YES  & YES & YES     &  no  & YES & YES    \\ \hline
Sequence Length    & YES  & no  & no   &  YES  & YES & YES    \\ \hline
Replicon           & YES  & no  & no     &  YES  & YES & YES    \\ \hline
Map Position       & YES  & YES & no     &  YES  & YES & no    \\ \hline
Product Mol Wt     & YES  & no  & no     &  no  & YES & no    \\ \hline
Product Subunits   & YES  & no  & no    &  no  & YES & no    \\ \hline
Product pI         & YES  & no  & no     &  no  & YES & no    \\ \hline
Product Ligands    & YES  & no  & no     &  no  & YES & no    \\ \hline
Evidence Code      & YES  & no  & no     &  no  & no  & no  \\ \hline
Cell Component     & YES  & no  & no     &  no  &  no & no    \\ \hline
GO Terms           & YES  & no  & YES     &  YES  & YES & YES    \\ \hline
Protein Features   & YES  & no  & YES     &  no  & YES & no    \\ \hline
Publication        & YES  & no  & no   &  YES  &  no & no    \\ \hline
Scaffold Length    & no  & YES  & no   &  YES & YES & no  \\ \hline
Scaffold GC Content  & no & no  & no   &  no & YES &  YES \\ \hline
Protein Family Assignment & no  & YES & YES  &  no & YES & YES  \\ \hline
Is Partial         & no  & no & no   &  no & YES & no   \\ \hline
Is Pseudogene      & YES & no & no   &  no & YES & YES  \\ \hline
%%  TOTALS           17     6    7      10     18     10
\end{tabular}
}
\caption{\label{tab:gene-protein-multi-search}
{\bf Gene/protein multi-search capabilities.}
Does the portal support multi-searches for genes and gene products based on
the data fields or criteria listed?  
``Publication'' means the ability to search for a gene based on a
publication cited in the pathway entry.
``Scaffold Length'' means the ability to search for a gene based on
the length of the scaffold it resides on.
``Protein Family Assignment'' means the ability to search for a gene
based on what protein families it is assigned to (e.g., Pfam or
TIGRFAM family).
``Is Partial'' means search for partial (truncated) proteins.
}
\end{table}

\begin{table}[!h]
\centerline{
\begin{tabular}{|l|c|c|c|c|c|c|c|} \hline
{\bf Tool}         & {\bf BioCyc} & {\bf KEGG} & {\bf Ensembl Bacteria} & {\bf KBase} & {\bf IMG} & {\bf PATRIC} \\ \hline \hline
Site Type                  & YES   & no    & no     & no  & no  & no \\ \hline
-- Attenuators             & YES   & no    & no     & no  &  no & no    \\ \hline
-- Origin of Replication   & YES   & no    & no     & no  &  no & no    \\ \hline
-- Phage Attachment Sites  & YES   & no    & no     & no  &  no & no    \\ \hline
-- REP Elements            & YES   & no    & no     & no  &  no & no    \\ \hline
-- Promoters               & YES   & no    & no     & no  &  no & no    \\ \hline
-- Terminators             & YES   & no    & no     & no  &  no & no    \\ \hline
-- mRNA Binding Sites      & YES   & no    & no     & no  & YES & no    \\ \hline
-- Riboswitches            & YES   & no    & no     & no  & YES & no    \\ \hline
-- TF Binding Sites        & YES   & no    & no     & no  & no  & no    \\ \hline
-- Transcription Units     & YES   & no    & no     & no  & no  & no    \\ \hline
-- Transposons             & YES   & no    & no     & no  & no  & no    \\ \hline
Replicon                   & YES   & no    & no     & no  & YES & no    \\ \hline
Map Position               & YES   & no    & no     & no  & YES & no    \\ \hline
Site Regulator             & YES   & no    & no     & no  & no  & no    \\ \hline
Site Ligands               & YES   & no    & no     & no  & no  & no    \\ \hline
Evidence Code              & YES   & no    & no     & no  & no  & no    \\ \hline
CRISPR Arrays              & no    & no    & no     & no  & YES & no      \\ \hline
%%   TOTALS                 17       0        0       0      5     0
\end{tabular}
}
\caption{\label{tab:site-multi-search}
{\bf DNA/RNA Site Multi-Search Capabilities.}
Does the portal support multi-searches for DNA and RNA sites based on
the data fields or criteria listed?  For example, does the portal
support searches for sites by the type of site (e.g., for attenuators
versus transcription-factor binding sites), and by numeric constraints
on the genome position of the site? 
}
\end{table}

\begin{table}[!h]
\centerline{
\begin{tabular}{|l|c|c|c|c|c|c|c|} \hline
{\bf Tool} & {\bf BioCyc} & {\bf KEGG} & {\bf Ensembl Bacteria} & {\bf KBase} & {\bf IMG} & {\bf PATRIC} \\ \hline \hline
Metabolite Page                       & YES & YES & no   & no  & no & no  \\ \hline
Chemical Similarity Search            & no  & YES & no   & no  & no & no  \\ \hline
Glycan Similarity Search              & no  & YES & no   & no  & no & no  \\ \hline
Reaction Page                         & YES & YES & no   & no  & YES & no  \\ \hline
-- Reaction Atom Mappings             & YES & YES & no    & no  & no  & no  \\ \hline
Individual Pathway Diagram            & YES & YES & no  & YES  & YES & YES \\ \hline
-- Automatic Pathway Layout           & YES & no  & no    & no  & no  & no  \\ \hline
-- Paint Omics Data onto Pathway      & YES & YES & no   & no  & YES  & no  \\ \hline
-- Depict Enzyme Regulation           & YES & no  & no    & no  & no  & no  \\ \hline
-- Depict Genetic Regulation          & YES & no  & no    & no  & no  & no  \\ \hline
-- Depict Metabolite Structures       & YES & YES (Tooltip) & no  & no  & no  & no  \\ \hline
Multi-Pathway Diagram                 & YES & no  & no    & no  & no  & no  \\ \hline
Full Metabolic Network Diagram        & YES & YES & no    & no  & no & no \\ \hline
-- Zoomable Metabolic Network         & YES & YES & no    & no  & no  & no \\ \hline
-- Paint Omics Data onto Diagram      & YES & no  & no    & no  & no & no \\ \hline
-- Animated Omics Data Painting       & YES & no  & no    & no  & no  & no \\ \hline
-- Metabolic Poster                   & YES & no  & no    & no  & no  & no \\ \hline
-- Organism Comparison                & YES & no  & no    & no  & no  & no \\ \hline
Automated Metabolic Reconstruction & YES (Desktop)\footnotemark[1] & YES & no    & YES  & YES & YES  \\ \hline
Enrichment Analysis (Pathways)        & YES & no  & no    & no  & YES  & no  \\ \hline
Execute Metabolic Model               & YES & no  & no    & YES   & no  & YES  \\ \hline
-- Gene Knock-out Analysis            & YES & no  & no    & YES   & no  & YES  \\ \hline
Chokepoint Analysis                   & YES & no  & no    & no  & no & no  \\ \hline
Dead-End Metabolite Analysis          & YES & no  & no    & no  & no & no \\ \hline
Blocked-Reaction Analysis             & YES & no  & no    & YES & no &  no \\ \hline
Route Search Tool                     & YES & YES & no    & no  & no & no  \\ \hline
Path Prediction Tool                  & no  & YES & no    & no  & no & no  \\ \hline
Assign EC Number                      & no  & YES & no    & no  & no & no  \\ \hline
%%  TOTALS                              24    14      0      7     5    4
\end{tabular}
}
\caption{\label{tab:metabolic-tools}
{\bf Metabolic Tools Comparison.}
$^{1}$ The desktop version of the Pathway Tools software performs
automated metabolic reconstruction.
}
\end{table}

\end{landscape}

\begin{table}[!h]
\centerline{
\begin{tabular}{|l|c|c|c|c|c|c|c|} \hline
          {\bf Tool} & {\bf BioCyc} & {\bf KEGG} & {\bf Ensembl Bacteria} & {\bf KBase} & {\bf IMG} & {\bf PATRIC} \\ \hline \hline
Name                 & YES          &     YES    &          no     &     no      &     YES   & YES\footnotemark[1]     \\ \hline
Database Identifier  & YES          &     YES    &          no     &     no      &     YES   & YES\footnotemark[1] \\ \hline
Ontology             & YES          &     no     &          no      &     no      &     YES   & YES   \\ \hline
Monoisotopic Mass    & YES          &     no     &          no      &     no      &   partial & no    \\ \hline
Molecular Weight     & YES          &     no     &          no      &     no      &   partial & no    \\ \hline
Chemical Formula     & YES          &     no     &          no     &     no      &   partial & no    \\ \hline
Chemical Substructure& YES          &     YES    &          no      &     no      &   partial & no    \\ \hline
InChi String         & YES          &     no     &          no      &     no      &   partial & no    \\ \hline
InChi Key            & YES          &     no     &          no      &     no      &   partial & no    \\ \hline
%%      TOTALS          9                  3                0             0             6          3
\end{tabular}
}
\caption{\label{tab:compound-multi-search}
{\bf  Compound multi-search capabilities.  }
Does the portal support multi-searches for chemical compounds based on
the data fields or criteria listed?  ``Ontology'' means the ability to
search for compounds based on a chemical ontology (classification).
$^{1}$This search will find pages of antimicrobial compounds.
}
\end{table}

\begin{table}[!h]
\centerline{
\begin{tabular}{|l|c|c|c|c|c|c|c|} \hline
       {\bf Tool} & {\bf BioCyc} & {\bf KEGG} & {\bf Ensembl Bacteria} & {\bf KBase} & {\bf IMG} & {\bf PATRIC} \\ \hline \hline
Name              &        YES   &      YES   &      no         &      no     &      YES  &      YES     \\ \hline
Ontology          &        YES   &      YES   &      no          &      no     &      YES  &      YES    \\ \hline
Size in Reactions &        YES   &      no    &      no         &      no     &      no   &      no    \\ \hline
Substrates        &        YES   &      YES   &      no          &      no     &      YES  &      no    \\ \hline
Evidence Code     &        YES   &      no    &      no          &      no     &      no   &      no    \\ \hline
Publication       &        YES   &      no    &      no          &      no     &      no   &      no    \\ \hline
%% TOTALS                   6            3            0                 0            3             2
\end{tabular}
}
\caption{\label{tab:pathway-multi-search}
{\bf Pathway multi-search capabilities.}
Does the portal support multi-searches for pathways based on
the data fields or criteria listed?  ``Ontology'' means the ability to
search for pathways based on a pathway ontology (classification).
}
\end{table}

\begin{table}[!h]
\centerline{
\begin{tabular}{|l|c|c|c|c|c|c|c|} \hline
{\bf Table Capability}   & {\bf BioCyc} & {\bf KEGG} & {\bf Ensembl Bacteria} & {\bf KBase} & {\bf IMG} & {\bf PATRIC} \\ \hline \hline
Table Datatypes:         &  &   &   &   &  &  \\ \hline
\ \ \ Genomes                              &  no & no  & no  & no  & no & YES \\ \hline
\ \ \ Genes                                & YES & no  & no  & no  & YES  & YES\footnotemark[1] \\ \hline
\ \ \ Proteins                             & YES & no  & no  & no  & YES  & YES \\ \hline
\ \ \ RNAs                                 & YES & no  & no  & no  & YES  & YES \\ \hline
\ \ \ Metabolites                          & YES & no  & no  & no  & partial  & no  \\ \hline
\ \ \ Pathways                             & YES & no  & no  & no  & partial  & YES \\ \hline
\ \ \ Reactions                            & YES & no  & no  & no  & partial  & no \\ \hline
\ \ \ Promoters                            & YES & no  & no  & no  & no  & no \\ \hline
\ \ \ Terminators                          & YES & no  & no  & no  & no  & no \\ \hline
\ \ \ Transcription Factor Binding Sites   & YES & no  & no  & no  & no  & no  \\ \hline
\ \ \ Transcription Units                  & YES & no  & no  & no  & partial  & no \\ \hline
\ \ \ Publications                         & YES & no  & no  & no  & no  & no \\ \hline
\ \ \ Transciptomics Experiments           & no  & no  & no  & no  & partial & YES \\ \hline
\ \ \ Biosynthetic Clusters                & no  & no  & no  & no  & YES & no  \\ \hline
\ \ \ Protein Families                     & no  & no  & no  & no  & no  & YES \\ \hline
Create Table from Uploaded File            & YES & no  & no  & no  & YES & YES \\ \hline
Create Table from database query result    & YES & no  & no  & no  & YES & YES \\ \hline
Include Database Properties as Table Columns & YES & no & no & no  & YES & YES \\ \hline
Create Columns as Computational Transformations & YES & no   & no  & no  & no  & no  \\ \hline
Set Operations Among Tables                & YES & no  & no  & no  & YES & YES \\ \hline
Filter Table Rows                          & YES & no  & no  & no  & YES & YES \\ \hline
Export Table to File                       & YES & no  & no  & no  & YES & YES \\ \hline
Share Table with Selected Users            & YES & no  & no  & no  & YES & YES \\ \hline
Share Table to the Public                  & YES & no  & no  & no  & no  & YES \\ \hline
%%   TOTALS                                  20    0     0      0    13.5   15
\end{tabular}
}
\caption{\label{tab:smt}{\bf Table-Based Analysis Capabilities.}
$^{1}$PATRIC provides tables of genomes and tables of features (defined sections of a genome, e.g., genes, CDS, mRNAs). 
}
\end{table}

\begin{table}[!h]
\centerline{
\begin{tabular}{|l|c|c|c|c|c|c|c|} \hline
{\bf Feature} & {\bf BioCyc} & {\bf KEGG} & {\bf Ensembl Bacteria} & {\bf KBase} & {\bf IMG} & {\bf PATRIC} \\ \hline \hline
Gene Page Load Time (sec)\footnotemark[1]  & 4.4 & 2.5 & 10.0 & 9.8 & 13.5 & 34.9  \\ \hline
Tooltips                  & YES & no  & YES   & YES  &  YES & YES   \\ \hline
User Guide                & YES & YES & YES\footnotemark[2] & YES &  YES & YES   \\ \hline
Webinars                  & YES & no  & YES\footnotemark[2] & YES &  YES & YES   \\ \hline
Workshops                 & YES & ?   &    YES & YES  &  YES & YES   \\ \hline
%%%%Publications  & 4853 & 20008 & 431 & 48 & 2176 & 593 \\ \hline
\end{tabular}
}
\caption{\label{tab:user-exp}
{\bf User Experience Features}
}
$^{1}$The extent of gene details and visualization displayed is vastly
different among sites and can lead to longer page load times. 
$^{2}$Userguide and webinars cover multiple Ensembl portals, not specifically bacteria. 
\end{table}

\subsection{Genomics Tools}

Genomics tools enable researchers to query, analyze, and compare
genome-related information within an organism DB.
Table~\ref{tab:genomics-tools} assesses most genomics tools;
Tables~\ref{tab:gene-protein-multi-search} and
\ref{tab:site-multi-search} describe genomics multi-search tools.

An explanation of the rows within Table~\ref{tab:genomics-tools} is as follows.

\begin{itemize}

\item {\bf Genome Browser}: Can a user browse a
  chromosome at different zoom levels to see the genomic features present?

\begin{itemize}
\item Are {\bf operons, promoters, and transcription-factor binding
  sites} depicted in the genome browser?

\item Is the {\bf nucleotide sequence} depicted in the genome browser?

\item {\bf Customizable Tracks}: Can a user add additional tracks to
  the genome browser, which show user-supplied data?

\item {\bf Comparative, by Orthologs}: Can a user compare 
  chromosome regions from several genomes side-by-side, 
  with orthologous genes indicated?  

\item {\bf Genome Poster}: Can the portal generate a printable, detailed,
  wall-sized poster of the entire genome, e.g., one that depicts every
  gene in the genome?
\end{itemize}

\item {\bf Retrieve Gene Sequence}: Can a user retrieve the
  nucleotide sequence of a gene?

\item {\bf Retrieve Replicon Sequence}: Can a user retrieve the
  nucleotide sequence of a specified region of a replicon?

\item {\bf Retrieve Protein Sequence}: Can a user retrieve the
  amino-acid sequence of a protein?

\item {\bf Nucleotide Sequence Alignment Viewer}: Can a user compare the
  nucleotide sequence   of a gene with orthologs from other organisms?

\item {\bf Protein Sequence Alignment Viewer}: Can a user compare the
  amino-acid sequence   of a protein with orthologs from other organisms?

\item {\bf Protein Phylogenetic Tree Analysis}: Can a user construct a
  phylogenetic tree from a set of protein sequences?

\item {\bf Sequence Searching by BLAST}: Is searching for a sequence
  in a genome by BLAST supported?

\item {\bf Sequence Pattern Search}: Is sequence searching by short sequence patterns
  supported?

\item {\bf Sequence Cassette Search}: Is sequence searching by 
  protein family recognition patterns supported?

\item {\bf Orthologs}: Can a user query for the orthologs of a given gene in other
  organisms?

%%%%\item {\bf Gene Regulatory Network Browser}: Does the portal support graphical
%%%%  browsing of the gene regulatory network of a genome?
%%%%
\item {\bf Gene/Protein Page}: Does the portal provide gene pages,
  showing relevant information such as the gene products and links to
  other DBs?

\item {\bf Enrichment Analysis (GO Terms)}: Can a user find which
  GO terms are statistically enriched, given a set of genes?

\item {\bf Enrichment Analysis (Regulation)}: Given a set of genes,
can a user compute which regulators of those genes are statistically
over-represented in the gene set?

\item {\bf Omics Dashboard}: Can a user submit a transcriptomics
  dataset for analysis using a visual dashboard tool that enables
  interactive summarization and exploration of the dataset in a manner
  similar to the BioCyc Omics Dashboard \cite{DashGene17}?
  
\item {\bf Multi-Organism Comparative Analysis}: Can a user globally
  compare a variety of different data types between several organisms?

\item {\bf Horizontal Gene Transfer Prediction}: Can the site show
  which genes may have been acquired by horizontal gene transfer?

\item {\bf Fused Protein Prediction}: Can the portal show
which genes result from fusions of genes that can be found separately in other organisms?

\item {\bf Alternative ORF Search (6-frame translation)}: 
Can a user assess alternative ORFs to the ones predicted on a given genomic region? Change the name to Alternate ORF View?

\item {\bf Genome Multi-Search}: Does the portal support
search and retrieval across all genomes using sequencing, environmental, or other metadata attributes?

\item {\bf gANI (Whole-genome Average Nucleotide Identity) Computations}:
  Whole-genome based average nucleotide identity (gANI) has been
  proposed as a measure of genetic relatedness of a pair of genomes.
  gANI for a pair of genomes is calculated by averaging the nucleotide
  identities of orthologous genes. The fraction of orthologous genes
  (alignment fraction or AF) is also reported as a complementary
  measure of similarity of the two genomes.

\item {\bf Kmer Frequency Analysis}: Can the portal display principal component
  analysis plots of oligonucleotide frequencies along genome length;
  allow comparison of genomes by the similarity of oligonucleotide
  composition, and identify sequences with abnormal
  oligonucleotide composition, such as horizontally transferred
  sequences and contaminating contigs/scaffolds?

\item {\bf Synteny Comparisons}: Does the portal provide a 
tool for evaluating conservation of
  gene order by plotting pairwise genome alignment?
  Potential translocations, inversions, or gaps relative to
  reference can be visualized.  Such a tool gives a quick snapshot of how closely
  related two strains might be.

\item {\bf Proteome Comparisons}: Find proteins that are
  shared between two or more genomes or unique to a given genome.

\item {\bf Statistical Analysis, Genome}: 
Example statistical analyses include 
counts of genes assigned to a ``feature'' (such as presence of a
COG/Pfam/TIGRFAM/KEGG domains), and counts of genes in different
Gene Ontology categories.

\item {\bf Statistical Analysis, Expression}: 
Does the portal provide tools for calculating statistical significance
of gene expression data?

\item {\bf Genome Function Comparison}: Genomes can be clustered based on a function profile
(e.g., COG/Pfam/TIGRFAM/KEGG features) and viewed as a hierarchical
cluster tree, principal component analysis, principal coordinate analysis plot, or other options,
to assess relatedness of selected genomes.

\item {\bf Insert Genomes into Reference Trees}:
Enables a user to determine evolutionary relationships between a
genome of interest and nearby reference genomes by building a tree of 
49 concatenated universal sequences.

\item {\bf Predict Effects of Sequence Variants}:
Enables users to predict effects of variation, including SNPs and
indels on transcripts in the region of the variant.

\end{itemize}

\subsection{Metabolic Tools}

Metabolic tools enable researchers to query, analyze, and compare
information about metabolic pathways and reactions within an organism
DB, to run metabolic models, and to analyze high-throughput data
in the context of metabolic networks.
Table~\ref{tab:metabolic-tools} assesses most metabolic tools;
Table~\ref{tab:compound-multi-search} describes metabolite multi-search
capabilities and Table~\ref{tab:pathway-multi-search} describe pathway
multi-search capabilities.

An explanation of the rows within Table~\ref{tab:metabolic-tools} is as follows.

\begin{itemize}

\item {\bf Metabolite Page}: Does the site provide a metabolite page,
  showing relevant information such as synonyms, chemical structure,
  and reactions in which the metabolite occurs?

\item {\bf Chemical Similarity Search}: Can the user search for
  chemicals that have similar structures to a provided chemical?

\item {\bf Glycan Similarity Search}: Can the user search for
glycans that have similar structures to a provided glycan?

\item {\bf Reaction Page}: Does the site provide a reaction page,
  showing relevant information such as EC numbers, reaction equation,
  and enzymes catalyzing the reaction?

\item {\bf Reaction Atom Mappings}: Can the reaction equation
  be shown with metabolite structures that depict the trajectories of atoms
  from reactants to products?

\item {\bf Pathway Diagrams}: Can pathway diagrams be depicted?

\item {\bf Automatic Pathway Layout}: Are pathway diagrams generated
  automatically by the software, thereby avoiding manual drawing?

\item {\bf Paint Omics Data onto Pathway}: Can a user visualize 
   omics data on pathway diagrams?

\item {\bf Depict Enzyme Regulation}: Can pathway diagrams
   show regulation of enzymes by metabolites, to depict information such as feedback inhibition?

\item {\bf Depict Genetic Regulation}: Can pathway diagrams
   show genetic regulation of enzymes, such as by transcription
   factors and attenuation?

\item {\bf Depict Metabolite Structures}: Can pathway diagrams
  show the chemical structures of metabolites?

\item {\bf Multi-Pathway Diagram}: Can users interactively create
  diagrams consisting of multiple interacting metabolic pathways?

\item {\bf Full Metabolic Network Diagram}: Can the entire metabolic
  reaction network of a genome be depicted and explored by an interactive graphical
  interface?

\item {\bf Zoomable Metabolic Network}: Does the metabolic network
  browser enable zooming in and out?

\item {\bf Paint Omics Data onto Network}: Can a user visualize 
  an omics dataset (e.g., gene expression, metabolomics) on the
  metabolic network diagram?

\item {\bf Animated Omics Data Painting}: Can several omics data points be
  visualized as an animation on the metabolic network diagram?

\item {\bf Metabolic Poster}: Can the portal generate a printable
  wall-sized poster of the organism's metabolic network?

\item {\bf Organism Comparison}: Can a user compare the metabolic
  networks of two organisms via the full metabolic network diagram?
  
\item {\bf Automated Metabolic Reconstruction}: Starting from a
  functionally annotated genome, can the metabolic reaction network (and
  pathways) be inferred in an automated fashion?

\item {\bf Enrichment Analysis (Pathways)}: Can the site compute 
   statistical enrichment of pathways within a large-scale dataset?

\item {\bf Execute Metabolic Model}: Can a user execute a steady-state
  metabolic flux model via the portal?

\item {\bf Gene Knock-out Analysis}: Can a user run flux-balance analysis (FBA) on the metabolic
  network by systematically disabling (knocking-out) various genes, to investigate
  how knock-outs perturb the network, and to predict gene essentiality?

\item {\bf Chokepoint Analysis}: Can the site compute chokepoint
  reactions (possible drug targets) in the full metabolic reaction network?  A chokepoint
  reaction is a reaction that either uniquely consumes a specific
  reactant or uniquely produces a specific product in the metabolic
  network. 

\item {\bf Dead-End Metabolite Analysis}: Can the portal compute dead-end
  metabolites in the full metabolic reaction network?  Dead-end
  metabolites are those that are either only consumed, or only
  produced, by the reactions within a given cellular compartment,
  including transport reactions.

\item {\bf Blocked-Reaction Analysis}: Can the portal compute blocked
  reactions in the full metabolic reaction network?  Blocked reactions
  cannot carry flux because of dead-end metabolites upstream or
  downstream of the reactions.

%%%%\item {\bf Metabolite Tracing Tool}: Can a user navigate the metabolic
%%%%  reaction network in an interactive fashion, moving from a starting
%%%%  metabolite along selectable reactions to a series of further
%%%%  metabolites?

\item {\bf Route Search Tool}: Given a starting and an ending
  metabolite, can the site compute an optimal series
  of known reactions (routes) that converts the starting metabolite to the
  ending metabolite?

\item {\bf Path Prediction Tool}: Given a starting
  chemical compound, can the site predict a series of previously
  unknown enzyme-catalyzed reactions that will act upon the input
  compound and the products of previous reactions?

\item {\bf Assign EC Number}: Can the portal compute an appropriate
  Enzyme Commission number for a user-provided reaction?

\end{itemize}

\subsection{Regulation Tools}
\label{sec:regulation}

BioCyc has a number of regulatory informatics tools that are not
provided by any of the portals.  We list those tools here rather than
providing a table.

\begin{itemize}
\item BioCyc includes a network browser that depicts the full transcriptional regulatory network
of the organism.  The network diagram can be queried interactively and
painted with transcriptomics data.

\item The BioCyc transcription-unit page depicts operon structure 
including promoters, transcription factor binding sites, and terminators,
the evidence for each, and describes regulatory interactions between
these sites and associated transcription factors and small RNA regulators.

\item BioCyc generates diagrams that summarize all regulatory influences on a gene, including
regulation of transcription, translation, and of the gene product.

\item BioCyc depicts  transcription-factor regulons as diagrams of all operons
regulated by a transcription factor.

\item BioCyc can depict regulatory influences on metabolism by
highlighting the regulon of a transcription factor on the BioCyc metabolic
map diagram.

\item BioCyc SmartTables can list the regulators or regulatees of each
gene within a SmartTable.

\item BioCyc can generate a report comparing the regulatory networks of two or more organisms.

\end{itemize}

\subsection{Advanced Search and Analysis}

These tools (see Table~\ref{tab:advanced-tools}) enable researchers to perform complex searches and analyses, to
retrieve data via web services and bulk downloads, and to create and
manipulate user accounts.

\begin{table}[!h]
\centerline{
\begin{tabular}{|l|c|c|c|c|c|c|c|} \hline
{\bf Tool}              & {\bf BioCyc} & {\bf KEGG} & {\bf Ensembl Bacteria} & {\bf KBase} & {\bf IMG} & {\bf PATRIC} \\ \hline \hline
Advanced Search         & YES   & no  & no   & no       & YES  & no  \\ \hline
Cross-Organism Search   & YES   & YES & YES  & partial  & YES  & YES   \\ \hline
Web Services            & YES   & YES & YES  & YES      & no  & no   \\ \hline
Other Query Options     &  *    &  *  &  *   &  *       & *   & *    \\ \hline
User Account            & opt/req & no & optional & required & opt/req & opt/req \\ \hline
Custom Notifications    & YES     & no & no & no & no & no \\ \hline
\multirow{3}{*} \ Download Formats     & biopax,gff   & json,sbml   & fasta,gff,gff3 & genbank,gff,tsv &  fasta,txt    & csv,fasta,gff    \\
                          & genbank &    &  json,mysql,rdf & fasta,json,sbml  &  &  embl,json   \\
                          & sbml    &    &                 &                  &  &  genbank     \\ \hline
%%  TOTALS                     5      2         3               2.5         3            2
\end{tabular}
}
\caption{\label{tab:advanced-tools}
{\bf  Comparison of Advanced Search and Analysis, Web
Services, and User Accounts.}  ``Opt/Req'' means that user accounts
  are optional for some operations and required for other operations.
  IMG also provides for downloading of reads, assemblies, QC reports,
  annotations, and more.
}
\end{table}

An explanation of the rows within Table~\ref{tab:advanced-tools} is as follows.

\begin{itemize}

\item {\bf Advanced Search}: Does the site enable the user to
  construct multi-criteria queries that search arbitrary DB
  fields using combinations of AND, OR, and NOT?

\item {\bf Cross-Organism Search}: Can a user search all organisms, 
specified organism sets, or taxonomic groups of organisms, for genes, metabolites, or pathways?

\item {\bf Web Services}: Can  DBs within the portal be queried
  programmatically by means of web services, using for example XML
  protocols?

\item {\bf Other Query Options:} What other query options are provided
  by the portal?
  \begin{itemize}
    \item BioCyc supports queries via its BioVelo query language
      \cite{BioVeloLangURL}.  Users can download BioCyc data files for
      text searches, and can load those data files into SRI's
      BioWarehouse system for SQL query access.  Users can download
      bundled versions of subsets of BioCyc plus Pathway Tools, and
      query the DBs via APIs for Python, Lisp, Java, Perl, and R.

    \item Users can download KEGG data files for text searches.

    \item Ensembl Bacteria provides a Perl API and public MySQL servers.

    \item KBase includes “code cells” for adding python code blocks to
      enable custom analyses, for which applications do not exist, or
      for programmatically calling Kbase native apps to automate large
      scale analyses.

    \item PATRIC provides a downloadable command line
      interpreter application that allows interactive submission of
      DB queries using a query language.

  \end{itemize}

\item {\bf User Account}: Are user accounts available for logging in, and
  for storing data and preferences?  ``Opt/Req'' means accounts are
  optional for some operations and required for other operations.

\item {\bf Custom Notifications}: Does the portal enable the user to
  register to be notified of curation updates in biological areas of
  interest to the user?

\item {\bf Bulk Download Formats}: What formats are supported by the
  portal for large scale data downloads?

\end{itemize}

\newpage

\subsection{Table-Based Analysis Tools}

Table-based analysis tools
enable users to define lists of genes, proteins, metabolites, or
pathways that are stored within the portal, and can be displayed,
analyzed, manipulated, and shared with other users.
These tools are called SmartTables by BioCyc and are called Carts by IMG.
A typical series of SmartTable operations are to define a SmartTable
containing a list of genes (such as from a transcriptomics
experiment); to configure which DB properties are displayed for
each gene within the SmartTable (such as displaying the gene name,
accession number, product name, and genome map position); performing a
set operation on the SmartTable such as taking the intersection with
another gene SmartTable; and transforming the gene SmartTable to say
a SmartTable  of the metabolic pathways containing those genes, or the
set of transcriptional regulators for those genes.

KBase does not have a tables mechanism, but it does have a data
sharing mechanism called narratives, which is not table-based.

An explanation of the rows within Table~\ref{tab:smt} is as follows.

\begin{itemize}

\item {\bf Datatypes Tables can Contain}: What types of entities may
  be stored in tables within each portal?  The more types of entities
  can be manipulated within tables, the more versatile the table
  mechanism is.  

\item {\bf Create Table from Uploaded File}:
Can tables be defined by uploading a data file that lists the entities
within the table?

\item {\bf Create Table from DB Query Result}:  
Can tables be defined from the result of a query within the portal?

\item {\bf Include DB Properties as Table Columns}:  
Can a user add columns to the table containing information from the
DB about a given entity, such as the accession number of a gene
or the nucleotide coordinate of a gene, or a diagram of the chemical
structure of a metabolite?

\item {\bf Create Table Columns as Computational Transformations}:  
Can table columns contained information computed from another column,
such as adding a column that computes the pathways in which a gene participates?

\item {\bf Set Operations Among Tables}:  
Can the portal create a new table by computing set operations between
two other tables, such as taking the union of the list of genes in two
other tables?

\item {\bf Filter Table Rows}:  
Can the portal remove rows from a table according to a search, such as
removing all entries from a table of metabolites where the metabolite
name contains ``arginine''?

\item {\bf Export Table to File}:  
Can the portal export the contents of a table to a data file?

\item {\bf Share Table with Selected Users}:  
Can a user share a table with a specific set of users?

\item {\bf Share Table with the Public}:  
Can a user share a table with the general public?

\end{itemize}

\newpage

\subsection{Data Content among the Portals}

\begin{table}[!h]
\centerline{
\begin{tabular}{|l|c|c|c|c|c|c|c|} \hline
{\bf Data Type}                     & {\bf BioCyc} & {\bf KEGG} & {\bf Ensembl Bacteria} & {\bf KBase} & {\bf IMG} & {\bf PATRIC} \\ \hline \hline
Genome Count                        & 14,560       & 5,130      & 44,046               & 122,688      & 97,179   & 184,000 \\ \hline
\ \ \ Bacterial Genomes             & 14,134       & 4,854      & 43,552               & 121,994      & 66,362   & 181,260 \\ \hline
\ \ \ Archaeal Genomes              &  394         & 276        & 494                  & 694          & 1,724    & 2,881   \\ \hline
\ \ \ Uncultivated Organisms        &     &     &  0       &  & 11,466 & 0    \\ \hline
Genome Metadata                     & YES & YES & no   & no  & YES      & YES \\ \hline
Regulatory Networks                 & 11  & no  & no   & no  & no       & no  \\ \hline
Protein Localization                & YES & no  & no   & no  & no       & no  \\ \hline
Protein Features                    & YES & no  & YES  & no  & partial  & YES \\ \hline
Protein 3-D Structures              & no  & YES & no   & no  & no       & no \\ \hline
GO Terms                            & YES & no  & YES  & YES & YES      & YES \\ \hline
Evidence Codes                      & YES & no  & no   & no  & YES      & partial\footnotemark[1] \\ \hline
Operons                             & YES & no  & no   & no  & no       & YES  \\ \hline
Prophages                           & YES & no  & no   & no  & YES      & YES  \\ \hline
Growth Media                        & YES & no  & no   & YES & no       & no  \\ \hline
Gene Essentiality                   & YES & no  & no   & no  & no       & YES  \\ \hline
Gene Clusters for Secondary Metabolites& no & no & no  & no  &  YES     & no   \\ \hline
Gene Pairs with Correlated Expression &no & no  & no   & no  & no       & YES \\ \hline
Protein-Protein Interactions        & no  & no  & no   & no  & no       & YES \\ \hline
AMR Phenotypes                      & no  & no  & no   & no  & no       & YES \\ \hline
%% Total                              10    2     2      2     5.5         9.5
\end{tabular}
}
\caption{\label{tab:data-content}
{\bf Data Types Comparison.}
$^{1}$PATRIC includes evidence codes in only two DB tables. 
}
\end{table}

Table~\ref{tab:data-content} describes the types and quantities of
data present in each web portal.
An explanation of the rows within the Table~\ref{tab:data-content} is
as follows.  

\begin{itemize}

\item {\bf Genome Count (Bact./Arch.)}: How many bacterial genomes (organisms)
  does the portal provide access to?  Only bacteria and
  archaea are counted here, although some resources provide 
  eukaryotic and viral genomes.

\item {\bf Genome Metadata}: Does the portal contain genome metadata, such as the
  lifestyle of the organism, and the location of where the organism sample was
  obtained?

\item {\bf Regulatory Networks}: Is (gene) regulatory information provided by
  the site?  Eleven BioCyc DBs provide regulatory networks
    larger than 100 transcriptional regulatory interactions.

\item {\bf Protein Localization}: Does the portal contain protein cellular locations?

\item {\bf Protein Features}: Are annotations of features of protein
  sequences provided by the portal?  Such features include which
  residues bind to cofactors or to metal ions, and where signaling
  peptide sequences reside.  IMG provides transmembrane and signal
  peptide features.

\item {\bf GO Terms}: Are GO term annotations provided by the site?
  IMG provides evidence codes for GO terms.  BioCyc provides evidence
  terms for gene functions, pathway presence, operon presence.

\item {\bf Evidence Codes}: Are evidence codes for the annotations
  provided by the resource, so the level of validity of the data can
  be assessed?

\item {\bf Operons}: Are genes grouped into operons, where applicable?

\item {\bf Prophages}: Are potential prophages indicated on the
  genomes?

\item {\bf Growth Media}: Are growth media for known growth conditions
  of the organisms provided by the site?  (BioCyc provides
  growth-media data for two organisms.)

\item {\bf Gene Essentiality}: Are gene essentiality data under
  various growth conditions provided by the site?  (BioCyc provides
  gene-essentiality data for 36 organisms.)

\item {\bf Gene Clusters for Secondary Metabolites}: 
Does the site identify putative operons of genes encoding enzymes for the production of 
secondary metabolites?

\item {\bf Gene pairs with correlated expression}: Pairs of genes with correlated 
expression based on experimental evidence. 

\item {\bf Protein-Protein interactions}: Pairs of protein with either experimental or computational
evidence of interacting. 

\item {\bf AMR phenotypes}: Can the site display phenotypes for antimicrobial resistance (e.g., is a strain resistant 
or susceptible to a particular antimicrobial compound)? 

\end{itemize}

\subsection{User Experience}

Table~\ref{tab:user-exp} contains several features that reflect the usability of the various portals.  These include
average loading times for typical gene pages for each portal; 
%%%%the number of publications about the portal, 
and other features and resources that assist the user in learning to use each portal.
 
\begin{itemize}

\item {\bf Mean Load Time for Gene Pages}: Since gene pages are among the most
  commonly visited information pages within a genome web portal, the
  time required for the page to load in a web browser is central to
  the user experience.  The values in this row are the average
  number of seconds required for each portal to load a gene page.  The
  values are averaged across six sessions, conducted from Menlo Park,
  California and Richmond, Virginia to average out geographic
  distances to each portal. Each session tested five genes on
  each of the six portals.  Testing was conducted using the Chrome browser version
  68.0, running on MacOS 10.13.6.  Testing consisted of clearing the
  browser cache, and pasting the URL of the gene page into the
  browser.  The load was monitored using the 'Network' panel of
  Chrome's Developer Tools (More Tools $\rightarrow$ Developer Tools).  The page
  was allowed to completely load (including loading large files and
  waiting for Ajax calls to complete).  The number used is the ``Finish'' time in
  the bottom line of the panel.  While some portals were disadvantaged
  by starting from an empty cache, forcing large files to be loaded,
  others were slowed by long Ajax calls.  We have removed the single worst time recorded
  of the 30 times (5 genes x 6 sessions) for each portal.

\item{\bf Portal Information}: Lists the availability of a
  userguide, extensive explanatory tooltips throughout the site, recorded webinars (either downloadable
  files or on YouTube or similar site), and user workshops.

%%%%\item{\bf Publications}: The count of the number of citations, based
%%%%  on Web of Science for papers specifically describing the portal,
%%%%  rather than, for example, associated algorithms.  These counts were
%%%%  measured Aug 31, 2018.

\end{itemize}

\section{Discussion}

Table~\ref{tab:tallies} summarizes the number of capabilities present
in each portal.  In each row of Table~\ref{tab:tallies} we have summed the
counts in the column for each portal from the specified tables, with
each ``YES'' counted as 1, each ``partial'' counted as $1/2$, and each
``no'' counted as 0.

\begin{table}[!h]
\centerline{
\begin{tabular}{|l|c|c|c|c|c|c|c|} \hline
{\bf Tool}       &  {\bf BioCyc} & {\bf KEGG} & {\bf Ensembl Bacteria} & {\bf KBase} & {\bf IMG} & {\bf PATRIC} \\ \hline \hline
Major            &       51     &      30     &            14          &     27.5    &     35     &     29        \\ \hline
SmartTables      &       20     &       0     &            0           &      0      &     13.5   &     15       \\ \hline
Multi-Search     &       49     &      12     &            7           &     10      &     32     &     15        \\ \hline
Data Types       &       10     &       2     &            2           &      2      &     5.5    &     9.5      \\ \hline
\end{tabular}
}
\caption{\label{tab:tallies}
{\bf Tallies of Portal Capabilities from Previous Tables.}
Row ``Major'' summarizes the major capabilities for
genomics tools, metabolic tools, and advanced tools 
present in Tables~\ref{tab:genomics-tools}, \ref{tab:metabolic-tools},
and \ref{tab:advanced-tools}.  Row ``SmartTables'' summarizes the number of SmartTables capabilities
for each portal present in Table~\ref{tab:smt}.
Row ``Multi-Search'' summarizes the number of multi-search capabilities
for each portal present in Tables~\ref{tab:gene-protein-multi-search},
\ref{tab:site-multi-search}, \ref{tab:compound-multi-search}, and \ref{tab:pathway-multi-search}.
Row ``Data Types'' summarizes the number of datatypes provided by
each portal present in Table~\ref{tab:data-content}, from row ``Genome
Metadata'' downward.
}
\end{table}

BioCyc received the highest count (51) of major capabilities (which
does not count its unique regulatory capabilities listed in Section~\ref{sec:regulation}).  
IMG ranked second  with a count of 35.
KEGG, PATRIC, and KBase ranked third, fourth, and fifth with counts of
30, 29, and 27.5, respectively.  Ensembl Bacteria ranked sixth with a count of 14.

BioCyc has the most extensive multi-search capabilities, with IMG in
second place; these portals provide users with the most extensive
capabilities for finding desired information.  

IMG has the most genomics capabilities, with PATRIC and BioCyc second
and third.  Ensembl Bacteria
has the fewest genomics capabilities.  BioCyc and IMG have the most
powerful gene/protein multi-search capabilities.  BioCyc has the most
extensive capabilities for DNA/RNA site multi-searches.

BioCyc has the most extensive metabolic capabilities.  KEGG ranks
second; it lacks metabolic modeling capabilities, and it lacks
network analysis tools such as dead-end metabolite analysis and chokepoint
analysis.
BioCyc has the most extensive metabolic multi-search
capabilities, with IMG second.

SmartTables make extensive data analysis capabilities
available to users that in many cases would otherwise require
assistance from a programmer.  
BioCyc has the most extensive SmartTable capabilities, with PATRIC
ranking second and IMG ranking third.  KEGG, Ensembl Bacteria, and KBase completely lack
SmartTables capabilities.

PATRIC has the largest number of genomes, with KBase and IMB ranked
second and third, respectively;
KEGG has the smallest number
of genomes.  Most of the PATRIC genomes were assembled from
whole-genome shotgun data and thus are expected to be of lower
quality --- only 11,803 PATRIC bacterial genomes are complete genomes.

KEGG provides the fastest loading gene pages; BioCyc pages are the
second fastest.  Pages for KBase, Ensembl Bacteria, and IMG are significantly slower.  
PATRIC gene pages are the slowest, loading 13.96 times slower than KEGG gene pages.  

BioCyc contains the most extensive  analysis capabilities for
metabolomics and transcriptomics data, 
including painting omics data onto individual pathways, multi-pathway
diagrams, and zoomable metabolic maps; enrichment analysis for GO
terms, regulation, and pathways; and an Omics Dashboard.

BioCyc contains extensive unique content not included in any of the
other portals including regulatory network data, data on growth under
different nutrient conditions, 
experimental gene essentiality data, reaction atom mappings (also
present in KEGG), and thousands of textbook page
equivalents of mini-review summaries.
KEGG is particularly lacking a diverse range of datatypes, for
example, KEGG lacks protein features, localization information,
GO terms, and evidence codes.

\section{Conclusions}

Microbial genome web portals have a broad range of capabilities, and
are quite variable in terms of what capabilities they provide.  
We assessed the capabilities of BioCyc, KEGG, Ensembl Bacteria, KBase,
IMG, and PATRIC.
BioCyc provided the most capabilities overall in terms of
bioinformatics tools and breadth of data content; it also provides a
level of curated data content (curated from 89,000 publications) that far exceeds that within the other sites.  
IMG ranked second overall, second in bioinformatics tools, and second in number of genomes.
KEGG ranked third overall, PATRIC ranked fourth, KBase ranked fifth,
and Ensembl Bacteria ranked sixth.
IMG provided the most extensive genome-related tools, with BioCyc a
close second.
BioCyc provided the most extensive metabolic tools, with KEGG ranked
second.  Ensembl Bacteria provided no metabolic tools.
PATRIC provided the largest number of genomes.
BioCyc provided extensive regulatory network tools (and data) that are
not present in any of the other portals.
BioCyc provided the most extensive SmartTable tools and the most
extensive omics data analysis tools.

\section*{Acknowledgments}

%%%%We thank Natalia Ivanova, Nikos Kyrpides, and Rekha Seshadri of the Joint Genome Institute for helpful discussion and comments.
We thank Dr. Nishadi De Silva of the European Bioinformatics Institute
for comments and corrections 
regarding Ensembl Bacteria.
We thank the KBase team for comments and corrections regarding KBase.
We thank Maulik Shukla of Argonne National Laboratory and the
University of Chicago and Rebecca Wattam of Virginia Tech for comments and corrections regarding PATRIC.
Research reported in this publication was supported by SRI
International and by the National
Institute Of General Medical Sciences of the National Institutes of
Health under Award Number 5R01GM080745. The content is solely the
responsibility of the authors and does not necessarily represent the
official views of the National Institutes of Health.
For JGI contributors, the work presented in this paper was supported
by the Director, Office of Science, Office of Biological and
Environmental Research, Life Sciences Division, U.S. Department of
Energy under Contract No. DE-AC02-05CH11231.

%%%%\bibliography{../../all}
\bibliographystyle{plain}

\end{document}